\documentclass[aps,prl,twocolumn,10pt]{revtex4-1}
\usepackage[latin9]{inputenc}
\usepackage{amsmath}
\usepackage{amssymb}
\usepackage{graphicx}

\begin{document}

\title{Reveal quantum correlation in complementary bases}

\author{Shengjun Wu$^{1,2}$\footnote{Correspondence to sjwu@nju.edu.cn}, Zhihao Ma$^{3,4}$, Zhihua Chen$^{5,6}$, and Sixia Yu$^{2,6}$}

\affiliation{$^1$Kuang Yaming Honors School, Nanjing Univeresity, Nanjing, Jiangsu 210093, China \\
$^2$Department of Modern Physics and the Collaborative Innovation Center for Quantum Information and Quantum Frontiers,
University of Science and Technology of China, Hefei, Anhui 230026, China\\
$^3$Department of Mathematics, Shanghai Jiaotong University, Shanghai 200240, China \\
$^4$Department of Physics and Astronomy, University College London, WC1E 6BT London, United Kingdom \\
$^5$Department of Science, Zhijiang College, Zhejiang University of Technology, Hangzhou 310024, China \\
$^6$Centre for Quantum Technologies, National University of Singapore, 3 Science Drive 2, Singapore 117543, Singapore
}

\begin{abstract}
An essential feature of genuine quantum correlation is the simultaneous existence of correlation in complementary bases. We reveal this feature of quantum correlation by defining measures based on invariance under a basis change. For a bipartite quantum state, the classical correlation is the maximal correlation present in a certain optimum basis, while the quantum correlation is characterized as a series of residual correlations in the mutually unbiased bases. Compared with other approaches to quantify quantum correlation, our approach gives information-theoretical measures that directly reflect the essential feature of quantum correlation.
\end{abstract}

\maketitle

%\keywords{Quantum correlation | Mutually unbiased bases | Holevo quantity}

\iffalse
{\bf Significance Statement:}
Classical correlation that exists in a certain basis vanishes in any complementary bases, while genuine quantum correlation exists simultaneously in the complementary bases. This property of quantum correlation enables a peculiar feature called the ``spooky action at a distance" by Einstein.
Here, we present a natural way to quantify the ``spooky action at a distance", and quantitatively reveal the essential mystery of genuine quantum correlation: the simultaneous existence in complementary bases.
\fi

%\section{Introduction}

\bigskip

\noindent
{\large \bf Introduction}

Quantum physics differs significantly from classical physics in many aspects. A complete
classical description of an object contains information concerning
only compatible properties, while a complete quantum description of an object also contains
complementary information concerning incompatible properties (see Fig. 1).
This difference is also present in correlations. A classical correlation
in a bipartite system involves the correlation of only a certain property,
while a quantum correlation in a bipartite system also involves complementary
correlations of incompatible properties. The simultaneous existence
of complementary correlations together with the freedom to select
which one to extract is the most important feature of quantum correlation (see Fig. 2).
Schr\"odinger introduced the word ``entanglement'' to describe this
peculiar feature, which was termed ``spooky action at a distance''
by Einstein \cite{EPR,Sc,ScBorn,EinsteintoBornletter}.

\begin{figure}[h]
\centering \includegraphics[width=8.5cm]{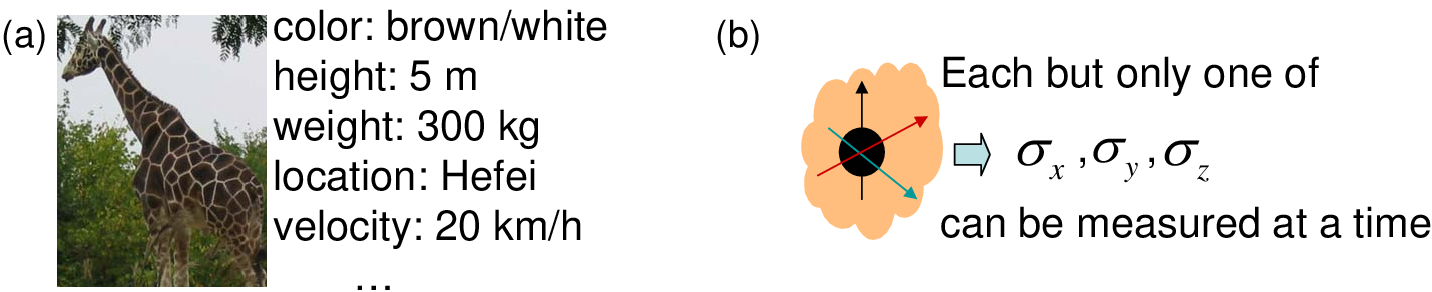} \caption{(a) A complete classical description of a classical object (a giraffe) is a simple collection
of information about compatible properties, such as color, height,
weight, position and velocity (the photo was taken by S.W. in Hefei animal zoo). (b) A complete quantum description
(a quantum state $\left| \psi \right\rangle$) of a quantum system (e.g. a spin-1/2 particle) contains information about incompatible properties ($\sigma_x$, $\sigma_y$, $\sigma_z$)
in an intrinsic way: information about incompatible properties exists
simultaneously even though only a single property can be measured at a time; and we can freely select which property
to measure.}
\label{figa}
\end{figure}

More recently, entangled states were defined as states that cannot be
written as convex sums of product states. This precise definition
is very helpful in terms of both mathematical and physical convenience,
and it motivates the useful definition of the entanglement of formation.
However, we now know that entanglement of formation
is just one particular aspect of quantum correlation.
Many measures of quantum correlation have been proposed from different perspectives, and they can be divided
into two categories: entanglement measures \cite{BBPS96,Bennett96hash,VPRK97,CW04,horodecki09rev},
and measures of nonclassical correlation
beyond entanglement \cite{OZ01,Piani2008,wu2009correl,Oppenheim2002,Horodecki2003a,Luo2008,Modi2010,Dakic2010,LuoFu2010,AD2010,GP2010,Modi2011,Alexander2013,Coles12,AFCA13}.

The essential feature of quantum correlation, i.e., the simultaneous existence of complementary correlations in different bases, is also revealed by the Bell's inequalities \cite{Bellinequality,Genovese2005}.
Bell's inequalities quantify quantum correlation via expectation values of local complementary observables. Instead, we shall seek a way to directly reveal the essential feature of quantum correlation from an information-theoretical perspective.
Indeed, there are several previous entropic measures of quantum correlation (such as quantum discord $D$, measurement-induced disturbance, symmetric discord, etc), which are proposed from an information-theoretical perspective. But these measures are based
on the difference between quantum mutual information \cite{GPW2005} (which is
assumed as the total correlation) and a certain measure of classical correlation.
Here, we take a different approach and reveal the essential feature of quantum correlation directly.
The genuine quantum correlation does not vanish under a change of basis,
and can be characterized as the residual correlations remaining in the
complementary bases.

\begin{figure}[h]
\centering \includegraphics[width=8cm]{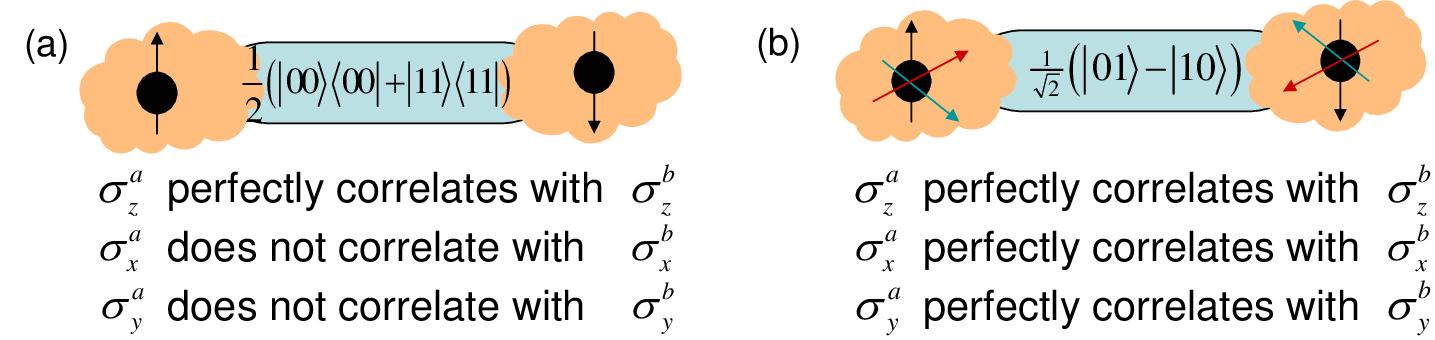}
\caption{
(a) Classical
correlation in a bipartite state reaches the maximum in a certain
basis and vanishes in any complementary basis. (b) However, quantum correlation
in a bipartite state contains correlations in complementary bases
simultaneously; and one can freely select with which basis to read out
the correlation.}
\label{figb}
\end{figure}

%\section{Classical and genuine quantum correlations}

\bigskip

\noindent
{\large \bf Results}

\noindent
{\bf The idea. }
We begin with a comparison between  the correlations in two different states:
\begin{eqnarray}
\rho_c &=& \frac{1}{2} (\left|00 \right\rangle \left\langle 00 \right| + \left|11 \right\rangle \left\langle 11 \right|) ,  \\
\left|EPR\right\rangle &=& \frac{1}{\sqrt{2}}\left(\left|01\right\rangle -\left|10\right\rangle \right).
\end{eqnarray}
The first state has only classical correlation, which can be revealed when Alice and Bob each measure the observable $\sigma_z$, i.e., project their qubits onto the basis $\left\{ \left|0\right\rangle ,\left|1\right\rangle \right\}$.  If they measure a complementary observable, $\sigma_x$ or $\sigma_y$, no correlation between their measurement results exists.
The second state is the Einstein-Podolsky-Rosen (EPR) state (or the singlet state), which has both classical and quantum correlations.
The classical correlation in the EPR state can be revealed when Alice and Bob each measure the same observable, e.g. $\sigma_z$.
Moreover, this kind of correlation also exists simultaneously in complementary bases (actually, in all bases).
The simultaneous existence of correlation in complementary bases is an essential feature of the genuine quantum correlation.
This feature is illustrated in Fig. 2 and treated in a rigorous manner in the rest of this article.

\medskip

\noindent
{\bf Classical and genuine quantum correlations. }
For any bipartite quantum state $\rho_{AB}$, there are many measures
of classical correlation \cite{wucorr2012}. Here, we use the one proposed
by Henderson and Vedral \cite{HV2001}, which is also used in the
definition of quantum discord \cite{OZ01}. Alice selects a basis
$\left\{ \left|a_{i}\right\rangle _{A}|i=1,\cdots,d_{A}\right\} $
of her system in a $d_{A}$-dimensional Hilbert space and performs
a measurement projecting her system onto the basis states. With probability
$p_{i}=tr_{AB}((\left|a_{i}\right\rangle _{A}\left\langle a_{i}\right|\otimes I_{B})\rho_{AB})$,
Alice will obtain the $i$-th basis state $\left|a_{i}\right\rangle $,
and Bob's system will be left in the corresponding state $\rho_{i}^{B}=_{A}\left\langle a_{i}\right|\rho_{AB}\left|a_{i}\right\rangle _{A}/p_{i}$.
The Holevo quantity of the ensemble $\{p_{i};\rho_{i}^{B}\}$ that is
prepared for Bob by Alice via her local measurement is given by $\chi\{\rho_{AB}|\{\left|a_{i}\right\rangle _{A}\}\}=\chi\{p_{i};\rho_{i}^{B}\}\equiv S(\sum_{i}p_{i}\rho_{i}^{B})-\sum_{i}p_{i}S(\rho_{i}^{B})$, which
denotes the upper bound of Bob's accessible information about
Alice's measurement result when Alice projects her system onto the
basis $\left\{ \left|a_{i}\right\rangle _{A}\right\} $. The classical
correlation in the state $\rho_{AB}$ is defined as the maximal Holevo
quantity over all local projective measurements on Alice's system:
\begin{equation}
C_{1}(\rho_{AB}) \equiv \max_{\{\left|a_{i}\right\rangle _{A}\}}\chi\{\rho_{AB}|\{\left|a_{i}\right\rangle _{A}\}\}.\label{defC1}
\end{equation}
A basis $\{\left|a_{i}\right\rangle _{A}\}$ that achieves the maximum $C_{1}(\rho_{AB})$ is called a $C_1$-basis of $\rho_{AB}$, and is denoted as  $\left\{ \left|\mathcal{A}_{i}^{1}\right\rangle _{A}|i=1,\cdots,d_{A}\right\} $. There could exist many $C_1$-bases for a state $\rho_{AB}$.

We consider another basis $\left\{ \left|a_{j}^{2}\right\rangle _{A}|j=1,\cdots,d_{A}\right\} $,
which is mutually unbiased to the $C_1$-basis $\left\{ \left|\mathcal{A}_{i}^{1}\right\rangle _{A}|i=1,\cdots,d_{A}\right\} $
in the sense that $\left|\left\langle \mathcal{A}_{i}^{1}|a_{j}^{2}\right\rangle \right|=\frac{1}{\sqrt{d_{A}}}$,
i.e., if the system is in a state of one basis, a projective
measurement onto the mutually unbiased basis (MUB) will yield each
basis state with the same probability. The most essential feature
of quantum correlation is that when Alice performs a measurement in
another basis $\left\{ \left|a_{j}^{2}\right\rangle _{A}|j=1,\cdots,d_{A}\right\} $
that is mutually unbiased to the $C_1$ basis, Bob's accessible information
about Alice's results, characterized by the Holevo quantity, does not
vanish. This residual correlation represents genuine quantum correlation
and can be used as a measure of the quantum correlation. Formally,
a measure of quantum correlation $Q_{2}(\rho_{AB})$ in the
state $\rho_{AB}$ is defined as the Holevo quantity of Bob's accessible
information about Alice's results, maximized over Alice's projective measurements in the bases that are mutually
unbiased to a $C_1$-basis $\left\{ \left|\mathcal{A}_{i}^{1}\right\rangle _{A}|i=1,\cdots,d_{A}\right\} $, and further maximized over all possible
$\left\{ \left|\mathcal{A}_{i}^{1}\right\rangle _{A}\right\} $ (if not unique),
i.e.,
\begin{equation}
Q_{2}(\rho_{AB}) \equiv \max_{\{\left|\mathcal{A}_{i}^{1}\right\rangle _{A}\}} \max_{\{\left|a_{j}^{2}\right\rangle _{A}\}}\chi\{\rho_{AB}|\{\left|a_{j}^{2}\right\rangle _{A}\}\} . \label{defQ2}
\end{equation}
where $\left\{ \left|a_{j}^{2}\right\rangle _{A}|j=1,\cdots,d_{A}\right\} $
is any basis mutually unbiased to the basis $\left\{ \left|\mathcal{A}_{i}^{1}\right\rangle _{A}|i=1,\cdots,d_{A}\right\} $.
A basis $\{\left|a_{j}^{2}\right\rangle _{A}\}$ that achieves the maximum quantum correlation $Q_2$ in (\ref{defQ2}) is called a $Q_2$-basis, and is denoted as
$\left\{ \left|\mathcal{A}_{j}^{2}\right\rangle _{A}|j=1,\cdots,d_{A}\right\} $.
If there is only one $C_1$-basis, the second maximization over the $C_1$-bases $\left\{ \left|\mathcal{A}_{i}^{1}\right\rangle _{A}\right\} $ in (\ref{defQ2}) is not necessary.
If there is more than one $C_1$-basis, and not all of them achieve the maximum in (\ref{defQ2}), then we redefine the $C_1$-bases as those that also achieve the maximum in (\ref{defQ2}).
In other words, the bases (if any) that achieve the maximum in (\ref{defC1}) but do not achieve the maximum in (\ref{defQ2}) will not be considered as $\left\{ \left|\mathcal{A}_{i}^{1}\right\rangle _{A}\right\} $ any more.
After this redefinition, if $\left\{ \left|\mathcal{A}_{i}^{1}\right\rangle _{A}\right\} $  is still not unique, then $\left\{ \left|\mathcal{A}_{j}^{2}\right\rangle _{A}\right\} $ depends on the choice of $\left\{ \left|\mathcal{A}_{i}^{1}\right\rangle _{A}\right\} $.
It is also obvious that $\left\{ \left|\mathcal{A}_{j}^{2}\right\rangle _{A} \right\} $ is mutually unbiased to $\left\{ \left|\mathcal{A}_{i}^{1}\right\rangle _{A}\right\} $.

Similar to the case of characterizing entanglement, a single quantity is not
sufficient to describe the full property of quantum correlation because there
could be many types of quantum correlation. Following the
same line of reasoning, we can define the residual correlation in
a third MUB as
\begin{equation}
Q_{3}(\rho_{AB}) \equiv \max_{\{\left|\mathcal{A}_{i}^{1}\right\rangle _{A}\}} \max_{\{\left|\mathcal{A}_{j}^{2}\right\rangle _{A}\}}  \max_{\{\left|a_{k}^{3}\right\rangle _{A}\}}\chi\{\rho_{AB}|\{\left|a_{k}^{3}\right\rangle _{A}\}\},
\label{defQ3}
\end{equation}
where $\left\{ \left|a_{k}^{3}\right\rangle _{A}|k=1,\cdots,d_{A}\right\} $
is any basis mutually unbiased to both $\left\{ \left|\mathcal{A}_{i}^{1}\right\rangle _{A}\right\} $
and $\left\{ \left|\mathcal{A}_{j}^{2}\right\rangle _{A}\right\} $.
An optimum basis $\{\left|a_{k}^{3}\right\rangle _{A}\}$ to achieve the maximum in (\ref{defQ3}) is called a $Q_3$-basis, and is denoted
as $\left\{ \left|\mathcal{A}_{k}^{3}\right\rangle _{A}|k=1,\cdots,d_{A}\right\} $. Similarly, we redefine the $Q_2$-bases $\left\{ \left|\mathcal{A}_{j}^{2}\right\rangle _{A} \right\} $ as those that are optimum in both (\ref{defQ2}) and (\ref{defQ3}),
and further redefine the $C_1$-bases $\left\{ \left|\mathcal{A}_{i}^{1}\right\rangle _{A}\right\} $ as those that are optimum in (\ref{defC1}), (\ref{defQ2}) and (\ref{defQ3}).

Suppose in this manner that we can define $M$ quantities for the measures of correlation,
which are conveniently written as a single correlation vector
$\overrightarrow{C} \equiv (C_{1},Q_{2},Q_{3},\cdots,Q_{M})$
for the state $\rho_{AB}$. The number $M$ cannot be greater than
the number of MUBs that exist in the $d_{A}$-dimensional Hilbert
space. The first quantity $C_{1}$ denotes the maximal classical
correlation present in the state $\rho_{AB}$, which can be revealed
when Alice performs a measurement of her system in a $C_1$-basis $\left\{ \left|\mathcal{A}_{i}^{1}\right\rangle _{A}|i=1,\cdots,d_{A}\right\} $.
As classical correlation will vanish when measured in a mutually unbiased
basis, all of the other quantities describe
genuine quantum types of correlation. The second quantity $Q_{2}$
denotes the maximal genuine quantum correlation, and $\left\{ \left|\mathcal{A}_{j}^{2}\right\rangle _{A}|j=1,\cdots,d_{A}\right\} $
denotes an optimum basis to reveal this correlation. The third quantity
$Q_{3}$ denotes another type of genuine quantum correlation, and $\left\{ \left|\mathcal{A}_{k}^{3}\right\rangle _{A}|j=1,\cdots,d_{A}\right\} $
denotes a basis to reveal the second type of quantum correlation.

The splitting of the correlation vector as a single classical component ($C_1$) and several quantum components ($Q_2$, $\cdots$, $Q_M$) is not artificial, in fact, this splitting captures the essential difference between classical correlation and quantum correlation. The quantity $C_1$ represents the maximal amount of correlation available in a single basis ($\left\{ \left|\mathcal{A}_{i}^{1}\right\rangle _{A}\right\} $). The quantity $Q_2$ represents the maximal amount of correlation that is available not only in the first basis $\left\{ \left|\mathcal{A}_{i}^{1}\right\rangle _{A}\right\} $ but also in a second complementary basis $\left\{ \left|\mathcal{A}_{j}^{2}\right\rangle _{A}\right\}$ (we know $C_1 \geq Q_2 \geq Q_3 \geq \cdots$ from the definition of these quantities). And $Q_3$  represents the maximal amount of correlation available not only in $\left\{ \left|\mathcal{A}_{i}^{1}\right\rangle _{A}\right\} $ and $\left\{ \left|\mathcal{A}_{j}^{2}\right\rangle _{A}\right\} $, but also in a third MUB $\left\{ \left|\mathcal{A}_{k}^{3}\right\rangle _{A}\right\} $. The amount $Q_3$ of correlation exists in $3$ MUBs while the amount $Q_2$ of correlation may only exist in $2$ MUBs. Thus, $Q_3$ represents the amount of correlation with a higher level of quantumness than that of $Q_2$, and may have more practical advantages when $3$ MUBs are necessarily used (e.g. entanglement-based QKD via $6$-state protocol).

It should be pointed out that the maximum number of MUBs that exist in a $d_A$-dimensional Hilbert space is not known for the general case.
When $d_A$ is a power of a prime number, a full set of $d_A +1$ MUBs exists; for other cases, there may not exist $d_A +1$ MUBs. For example, when $d_A=6$, only $3$ MUBs have been found yet, while $3$ is much less than $d_A +1=7$. Many interesting works can be found on the existence of MUBs in the literature \cite{Ivanovic81,Ivanovic97,WF89,BBRV2002,DEBZ10}.
Since there exist at least $3$ MUBs for any integer $d_A \geq 2$, the quantities $C_1$, $Q_2$, and $Q_3$ are well-defined for any $d_A \geq 2$. In many cases, we are interested in combined systems of qubits with $d_A$ being a power of $2$, thus, $d_A +1 $ MUBs exist and quantities $C_1$, $Q_2$, $\cdots$, $Q_{d_A +1}$ are all well-defined.
For an arbitrary $d_A$-dimensional Hilbert space, we don't make assumptions about the maximal number of MUBs that exist, we only assume that $M$ MUBs are available, where $M$ is less or equal to the maximal number of MUBs that exist. In many cases, we only discuss the first $3$ elements ($C_1$, $Q_2$, and $Q_3$) of the correlation vector for simplicity.

\medskip

\noindent
{\bf Examples.}
Now, we shall calculate the correlation vector for several families of bipartite states, and see how these measures in terms of MUBs are
well justified as measures of classical and genuine quantum correlations.

For a bipartite
pure state written in the Schmidt basis,
$\left|\psi\right\rangle _{AB}=\sum_{i}\sqrt{\lambda_{i}}\left|a_{i}\right\rangle \left|b_{i}\right\rangle $,
the maximal classical correlation can be revealed when Alice performs
her measurement onto her Schmidt basis $\left\{ \left|a_{i}\right\rangle \right\} $;
thus, one immediately has $C_{1}=S(\rho_{B})=\sum_{i}-\lambda_{i}\log_{2}\lambda_{i}$.
If Alice chooses another basis $\left\{ \left|a_{i}'\right\rangle \right\} $,
whenever she obtains a particular measurement result, Bob will be left
with a pure state. Therefore, one can easily obtain the maximal true
quantum correlation $Q_{2}=S(\rho_{B})=C_{1}$; any other
basis will yield the same amount of quantum correlation. Therefore, the correlation
vector for a bipartite pure state $\left|\psi\right\rangle _{AB}$
is given as $\overrightarrow{C}=(S(\rho_{B}),S(\rho_{B}),\cdots,S(\rho_{B}))$.
The correlation vector exhibits a unique feature of the correlations
in a pure state: the classical correlation is equal to the
quantum correlation revealed in any basis, and both values are
equal to the von Neumann entropy of the reduced density matrix on either side, which is
the usual measure of entanglement in a pure state.

A classical-quantum (CQ) state is a bipartite state that can be written as
\begin{equation}
\rho^{cq}=\sum_{i}q_{i}\left|i\right\rangle \left\langle i\right|\otimes\sigma_{i},
\end{equation}
where $\left\{ q_{i}\right\} $ is a probability distribution, $\left\{ \left|i\right\rangle | i=0,1,\cdots,d_A -1 \right\} $
is a basis of system A in a $d_A$-dimensional Hilbert space, and $\left\{ \sigma_{i}\right\} $ is a set
of density matrices of system B. The maximal classical correlation
is revealed when Alice performs her measurement in the basis $\left\{ \left|i\right\rangle \right\}$ \cite{wu2009correl};
thus, the maximal classical correlation in the CQ state $\rho^{cq}$
is given by $C_{1}=\chi\left\{ q_{i};\sigma_{i}\right\} =S(\sum_{i}q_{i}\sigma_{i})-\sum_{i}q_{i}S(\sigma_{i})$.
To calculate the amount of quantum correlation, Alice projects
her system onto another basis $\left\{ \left|a_{j}^{2}\right\rangle \right\} $
that is mutually unbiased to the optimum basis $\left\{ \left|i\right\rangle \right\} $
for classical correlation. From $\left|\left\langle i|a_{j}^{2}\right\rangle \right|^{2}=\frac{1}{d_{A}}$,
we have
\begin{eqnarray}
\left\langle a_{j}^{2}\right|\rho^{cq}\left|a_{j}^{2}\right\rangle &=& \sum_{i}q_{i}\left\langle a_{j}^{2}\right|(\left|i\right\rangle \left\langle i\right|)\left|a_{j}^{2}\right\rangle \sigma_{i} \nonumber \\
&=& \frac{1}{d_{A}}\sum_{i}q_{i}\sigma_{i}=\frac{1}{d_{A}}\rho_{B}.
\end{eqnarray}
For each different result $j$ that Alice obtains, Bob is left with the
same state $\rho_{B}=\sum_{i}q_{i}\sigma_{i}$; thus, Bob's state has
no correlation with Alice's result, and we immediately have $Q_{2}=Q_{3}=\cdots=0$
according to the definitions of these quantities. Hence, for a CQ
state, the correlation vector is given as $\overrightarrow{C}=(C_{1},0,\cdots,0)$.
The only correlation present in a CQ state is the classical correlation, and
the quantum correlation in any MUB vanishes!

\begin{figure}[h]
\centering \includegraphics[width=8.5cm]{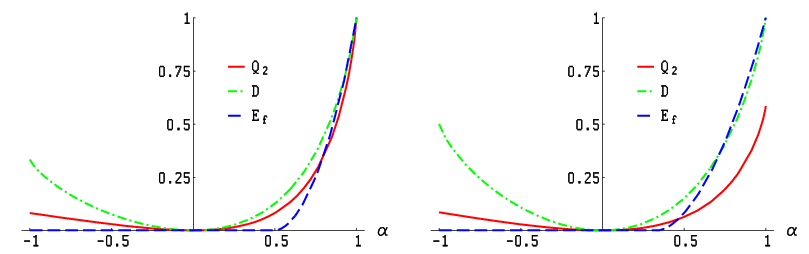}
\caption{Three measures of quantum correlation for the Werner states as functions
of $\alpha$ when $d=2$ (left) and $d=3$ (right). The red curve
represents our measure $Q_{2}$, the green curve represents
the quantum discord $D$ and the blue curve represents the entanglement
of formation $E_{f}$.}
\label{figc}
\end{figure}

Next, we consider the Werner states of a $d\times d$ dimensional
system \cite{Wer89},
\begin{equation}
\rho_{w}=\frac{1}{d(d-\alpha)}\left(I-\alpha P\right) ,
\label{Wernerstates}
\end{equation}
where $-1\leq\alpha\leq1$, $I$ is the identity operator in the $d^2$-dimensional Hilbert space, and $P=\sum_{i,j=1}^{d}\left|i\right\rangle \left\langle j\right|\otimes\left|j\right\rangle \left\langle i\right|$
is the operator that exchanges A and B. Because the Werner states are invariant
under a unitary transformation of the form $U\otimes U$, the maximal
classical correlation can be revealed when Alice simply projects her
system onto the basis states $\left\{ \left|i\right\rangle \right\} $.
With probability $p_{i}=\frac{1}{d}$, Alice will obtain the $i$-th
basis state $\left|i\right\rangle $, and Bob will be left with the
state $\rho_{i}^{B}=_{A}\left\langle i\right|\rho_{AB}\left|i\right\rangle _{A}/p_{i}=\frac{1}{d-\alpha}\left(I-\alpha\left|i\right\rangle \left\langle i\right|\right)$.
It is straightforward to show that $C_{1}=\chi\left\{ p_{i};\rho_{i}^{B}\right\} =\log_{2}(\frac{d}{d-\alpha})+\frac{1-\alpha}{d-\alpha}\log_{2}(1-\alpha)\equiv\chi_{w}$.
Due to the symmetry of the Werner states, it is not difficult to demonstrate that
$Q_{2}=Q_{3}=\cdots=C_{1}=\chi_{w}$. Therefore, for the Werner
state $\rho_w$, the correlation vector is given by $\overrightarrow{C}=(\chi_{w},\chi_{w},\cdots,\chi_{w})$.
The maximal quantum correlation in a Werner state can be revealed
in any basis, and it is equal to the maximal classical correlation
$C_{1}$. However, the correlation vector of a Werner state is different
from that of a pure state because $C_{1}\leq S(\rho_{B})=\log_{2}d$.
The inequality becomes an equality only when $d=2$ and $\alpha=1$, in which case,
the Werner state becomes a pure state $\rho_{w}=\left|EPR\right\rangle \left\langle EPR\right|$. For the Werner states, the symmetric discord
is equal to the quantum discord $D$ \cite{wu2009correl} when
Alice's measurement is restricted to projective measurements. The
entanglement of formation $E_{f}$ for the Werner states is given
as $E_{f}(\rho_{w})=h\left(\frac{1}{2}(1+\sqrt{1-[\max(0,\frac{d\alpha-1}{d-\alpha})]^{2}})\right)$,
with $h(x)\equiv - x\log_{2} x- (1-x) \log_{2}(1-x)$ \cite{wootters01}. The
three different measures of quantum correlation, i.e., our measure of maximal
quantum correlation $Q_{2}$, the quantum discord $D$ and
the entanglement of formation, are illustrated in Fig. 3
for comparison. From this figure, we see that the curve for entanglement of formation intersects the other two curves; thus, $E_f$ can be larger or smaller than $Q_2$ ($D$).

As the last example, we consider a family of two-qubit states, where the reduced density
matrices of both qubits are proportional to the identity operator.
Such a state can be written in terms of Pauli matrices,
\begin{equation}
\rho_{AB}=\frac{1}{4} (I_2\otimes I_2+\sum_{j,k=1}^{3}w_{jk}\sigma_{j}\otimes\sigma_{k} ),
\label{twoqubitsymmtricstates}
\end{equation}
where $I_2$ is the identity operator in the two-dimensional Hilbert space of a qubit, and $w_{jk}$ are real numbers that satisfy certain conditions to ensure the positivity of the matrix in (\ref{twoqubitsymmtricstates}). These two-qubit states can be transformed by a local unitary transformation
(that does not change the correlations)
to the following form:
\begin{equation}
\sigma_{AB}=\frac{1}{4} (I_2\otimes I_2+\sum_{j=1}^{3}r_{j}\sigma_{j}\otimes\sigma_{j} )
\label{twoqubitsymmtricstatessimplified}
\end{equation}
which is equivalent to the Bell-diagonal states.
To ensure the positivity of the matrix in (\ref{twoqubitsymmtricstatessimplified}), the real vector $\overrightarrow{r}=(r_1,r_2,r_3)$  must lie inside or on the boundary of the regular tetrahedron that is the convex hull of the four points:
$(-1,-1,-1)$, $(-1,1,1)$, $(1,-1,1)$ and $(1,1,-1)$ (which are the four Bell states).
The singular values of the matrix $w_{jk}$ are given by $|r_{j}|$.
We rearrange the three numbers $\{r_1,r_2,r_3\}$ according to their absolute values and denote the rearranged set as
$\{ \overline{r}_1, \overline{r}_2, \overline{r}_3\}$ such that $|\overline{r}_1| \geq |\overline{r}_2| \geq |\overline{r}_3|$.

\begin{figure}[h]
\centering \includegraphics[width=8.5cm]{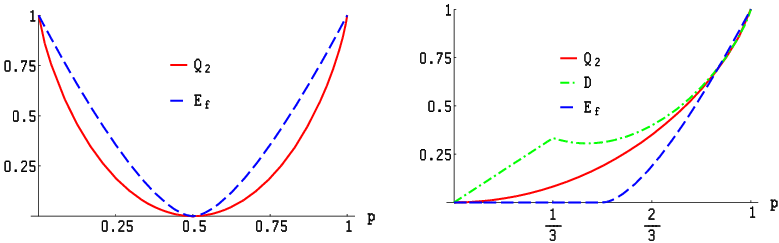}
\caption{Different measures of quantum correlation for two special classes of states:
$\rho_1 = p \left| \psi^- \right\rangle \left\langle \psi^- \right| +(1-p) \left| \psi^+ \right\rangle \left\langle \psi^+ \right|$ (left) and
$\rho_2 = p \left| \psi^- \right\rangle \left\langle \psi^- \right| +\frac{1-p}{2} \left( \left| \psi^+ \right\rangle \left\langle \psi^+ \right|+ \left| \phi^+ \right\rangle \left\langle \phi^+ \right| \right)$ (right).
In each figure, the red curve
represents our measure $Q_{2}$, the green curve represents
the quantum discord $D$, and the blue curve represents the entanglement
of formation $E_{f}$. In the left figure, the green curve is not shown because $D =Q_{2}$ for $\rho_1$.}
\label{figd}
\end{figure}

In the Methods, we prove that the correlation vector of the state in (\ref{twoqubitsymmtricstates}) is given by $\overrightarrow{C}=(\chi_{1},\chi_{2},\chi_{3})$,
where $\chi_{j}=1-h\left(\frac{1+\left| \overline{r}_{j} \right|}{2}\right)$ with $h(x)\equiv - x\log_{2}x- (1-x) \log_{2}(1-x)$.
To have some intuition of this result, we consider some special classes of states with only one parameter. When $r_1=r_2=r_3= - \frac{\alpha}{2-\alpha}$ with $-1 \leq \alpha \leq 1$, the states in (\ref{twoqubitsymmtricstatessimplified}) become the Werner states for $d=2$ in (6). When $r_1=r_2=1-2p$ and $r_3=-1$ with $0\leq p \leq 1$, the states in (\ref{twoqubitsymmtricstatessimplified}) become $\rho_1 = p \left| \psi^- \right\rangle \left\langle \psi^- \right| +(1-p) \left| \psi^+ \right\rangle \left\langle \psi^+ \right|$; we obtain $C_1 =1$ and $Q_2=Q_3=1-h(p)$. When $r_1=1-2p$ and $r_2 = r_3=-p$, the states in (\ref{twoqubitsymmtricstatessimplified}) become
$\rho_2 = p \left| \psi^- \right\rangle \left\langle \psi^- \right| +\frac{1-p}{2} \left( \left| \psi^+ \right\rangle \left\langle \psi^+ \right|+ \left| \phi^+ \right\rangle \left\langle \phi^+ \right| \right)$; we have $C_1=max\{1-h(p),1-h(\frac{1+p}{2})\}$, $Q_{2}=1-h(\frac{1+p}{2})$ and $Q_3=min \{1-h(p),1-h(\frac{1+p}{2})\}$. Here,
$\left|\psi^- \right\rangle = \left| EPR \right\rangle =\frac{1}{\sqrt{2}}(\left|01 \right\rangle - \left| 10 \right\rangle)$,
$\left|\psi^+ \right\rangle = \frac{1}{\sqrt{2}}(\left|01 \right\rangle + \left| 10 \right\rangle)$ and
$\left|\phi^+ \right\rangle = \frac{1}{\sqrt{2}}(\left|00 \right\rangle + \left| 11 \right\rangle)$.
Our measure of quantum correlation $Q_2$ is compared with the quantum discord $D$ and the entanglement of formation $E_f$ for $\rho_1$ and $\rho_2$ in Fig. 4.

%\section{Inequality relations between correlation measures}
\medskip

\noindent
{\bf Inequality relations between correlation measures.}
It is not difficult to show that the relation $Q_2 \leq D$ holds for the Werner states, and for all the example states considered in this article. However, it is not clear whether this inequality holds for any bipartite states. If $Q_2 \leq D$ holds for any bipartite states, then one can easily have $C_1 + Q_2 \leq S(A:B)$ where $S(A:B)=S(\rho_A)+S(\rho_B)-S(\rho_{AB})$ denotes the quantum mutual information.

Nevertheless, we can prove the following inequality:
\begin{equation}
C_1 + Q_2 \leq H_1 + H_2 + S(\rho_B) -S(\rho_{AB}) - \log_2 d_A  \label{C1plusQ2}
\end{equation}
where $H_{\gamma}$ ($\gamma =1,2$) denotes the Shannon entropy of the probability distribution $\{p_i^{(\gamma)}\}$ obtained by the measurement on system A in the basis $\left\{ \left|\mathcal{A}_{i}^{\gamma}\right\rangle _{A}|i=1,\cdots,d_{A}\right\} $. The proof is given in the Methods.
Since $H_{\gamma} \leq \log_2 d_A$, one immediately has
\begin{equation}
C_1 + Q_2 \leq S(\rho_B) -S(\rho_{AB}) + \log_2 d_A .  \label{C1plusQ2simple}
\end{equation}
As $C_1$ and $Q_2$ are the two largest elements in the correlation vector, when they are replaced by correlations in any two MUBs, inequalities (\ref{C1plusQ2}) and (\ref{C1plusQ2simple}) still hold.

%\section{Conclusions}
\bigskip

\noindent
{\large \bf Discussion}

\noindent
Our measures of quantum correlation provide a natural way to quantify the ``spooky action at a distance",
and directly reveal the essential feature of the genuine quantum correlation, i.e., the simultaneous existence of correlations in complementary bases.
This feature enables quantum key distribution with entangled states, since the quantum correlation that exists simultaneously in two ($k$) MUBs, which is quantified by $Q_2$ ($Q_k$), is the resource for entanglement-based QKD via two ($k$) MUBs. Quantitative relation between the genuine quantum correlation and the secret key fraction in QKD could be studied in further work.

All the measures considered above are not symmetric with respect to the exchange of systems A and B, as only system A's bases are considered to reveal the correlation. Symmetric measures and a symmetric correlation vector are also introduced and discussed in the Methods. A further study of the relation between the symmetric correlation vector and the symmetric discord \cite{wu2009correl} could reveal the difference between these measures in practical applications.

There are some open questions. Are our measures ($Q_2$, $Q_3$) of genuine quantum correlation additive?  How do they behave under some natural operations (for example, Alice adds an ancilla)?
Do our measures ($Q_2$, $Q_3$) behave like the entanglement measures that do not increase under local operations and classical communication (LOCC) \cite{horodecki09rev}, or more like the measures of nonclassical correlation beyond entanglement (for example, quantum discord) that could increase under LOCC \cite{SKB11}?
We hope that further investigations will unveil these mysteries.

%\section{acknowledgments}

%\section{Appendix}

%\subsection{Proof of the inequality (9)}

\bigskip

\noindent
{\large \bf Methods}

\noindent
{\bf Proof of the inequality (\ref{C1plusQ2}).}
Here we prove inequality (\ref{C1plusQ2}) in the main text.

Let $\{p_i^{(\gamma)}\}$  ($\gamma =1,2$) denote the probability distribution obtained by the measurement on system A in the basis $\left\{ \left|\mathcal{A}_{i}^{\gamma}\right\rangle _{A}|i=1,\cdots,d_{A}\right\} $, i.e.,
$p_i^{(\gamma)} = tr_{AB} (( \left|\mathcal{A}_{i}^{\gamma}\right\rangle \left\langle \mathcal{A}_{i}^{\gamma}\right| \otimes I)\rho_{AB})$. Let $H_{\gamma}$ ($\gamma =1,2$) denote the Shannon entropy of the probability distribution,
$H_{\gamma} = \sum_{i=1}^{d_A} - p_i^{(\gamma)} \log_2 p_i^{(\gamma)}$. Here, the basis $\left\{ \left|\mathcal{A}_{i}^{1}\right\rangle _{A}|i=1,\cdots,d_{A}\right\}$ is the optimum basis for measurement on system A to achieve the maximum classical correlation $C_1$, and the basis $\left\{ \left|\mathcal{A}_{i}^{2}\right\rangle _{A}\right\}$  is the optimum basis to achieve $Q_2$ among the bases that are mutually unbiased to $\left\{ \left|\mathcal{A}_{i}^{1}\right\rangle _{A}\right\}$. However, the proof below only requires that $\left\{ \left|\mathcal{A}_{i}^{1}\right\rangle _{A}\right\}$ and $\left\{ \left|\mathcal{A}_{i}^{2}\right\rangle _{A}\right\}$  are mutually unbiased to each other.

Let $\rho^{(\gamma)}= \sum_{i=1}^{d_A}  \left| \mathcal{A}_i^{\gamma} \right\rangle \left\langle \mathcal{A}_i^{\gamma} \right|  \otimes  p_i^{(\gamma)} \rho_{i}^{(\gamma)} $ with $\gamma=1,2$, where $\rho_{i}^{(\gamma)} =  \left\langle \mathcal{A}_i^{\gamma} \right| \rho_{AB} \left| \mathcal{A}_i^{\gamma} \right\rangle  / p_i^{(\gamma)}$.

The uncertainty relation \cite{BCCRRuncertainty} gives
\begin{equation}
S(\{\left|\mathcal{A}_{i}^{1}\right\rangle \} |B ) + S(\{\left|\mathcal{A}_{i}^{2}\right\rangle \} |B ) \geq \log_2 d_A + S(A|B)
\end{equation}
where $S(A|B) = S(\rho_{AB}) - S(\rho_B)$, and $S({\left|\mathcal{A}_{i}^{\gamma}\right\rangle } |B ) = S(\rho^{(\gamma)}) - S(\rho_B)$ ($\gamma =1, 2$).  As $\rho^{(\gamma)}$ is a CQ state, one can show that $S(\rho^{(\gamma)}) = H_{\gamma} +\sum_i p_i^{(\gamma)} S(\rho_i^{(\gamma)})$.
Therefore,
\begin{eqnarray}
H_1 + \sum_i p_i^{(1)} S(\rho_i^{(1)}) - S(\rho_B)  \nonumber \\
+ H_2 + \sum_i p_i^{(2)} S(\rho_i^{(2)}) - S(\rho_B)   \\
 \geq \log_2 d_A + S(\rho_{AB}) -S(\rho_B)   \nonumber
\end{eqnarray}
As $C_1 =  S(\rho_B) -\sum_i p_i^{(1)} S(\rho_i^{(1)}) $ and $Q_2 =S(\rho_B) -\sum_i p_i^{(2)} S(\rho_i^{(2)})$, we immediately have
\begin{equation}
C_1 + Q_2 \leq H_1 + H_2 + S(\rho_B) -S(\rho_{AB}) - \log_2 d_A
\end{equation}
which completes the proof of the inequality.

%\subsection{The correlation vector for a family of two-qubit states}
\medskip
\noindent
{\bf Calculation of the correlation vector for the states in (\ref{twoqubitsymmtricstates}).}
In this paragraph, we shall demonstrate that the correlation vector of the state in (\ref{twoqubitsymmtricstates}) is given by $\overrightarrow{C}=(\chi_{1},\chi_{2},\chi_{3})$,
where $\chi_{j}=1-h\left(\frac{1+\left| \overline{r}_{j} \right|}{2}\right)$ with $h(x)\equiv - x\log_{2}x- (1-x) \log_{2}(1-x)$.
We perform the calculation
in the transformed basis, with the states rewritten in Eq. (\ref{twoqubitsymmtricstatessimplified}).
Without loss of generality, we can suppose the numbers $r_j$ are already arranged according to $\left|r_{1}\right|\geq\left|r_{2}\right|\geq\left|r_{3}\right|$; then, we only need to prove that
$\overrightarrow{C}=(\chi_{1},\chi_{2},\chi_{3})$, where $\chi_{j}=1-h\left(\frac{1+\left| r_{j} \right|}{2}\right)$.
A projective measurement performed on qubit A can be written $P_{\pm}^{A}=\frac{1}{2}(I_2 \pm\overrightarrow{n}\cdot\overrightarrow{\sigma})$,
parameterized by the unit vector $\overrightarrow{n}$. We have
\begin{equation}
p_{\pm}\rho_{\pm}^{B}\equiv Tr_{A}\left(P_{\pm}^{A}\rho_{AB}\right)=\frac{1}{2}\cdot\frac{1}{2} (I_2 \pm\sum_{j}n_{j}r_{j}\sigma_{j}  ).
\end{equation}
When Alice obtains $\pm$, qubit B will be in the corresponding states
$\rho_{\pm}^{B}=\frac{1}{2}\left(I_2 \pm\sum_{j}n_{j}r_{j}\sigma_{j}\right)$,
each occurring with probability $\frac{1}{2}$. The entropy $S(\rho_{\pm}^{B})$
reaches its minimum value $h\left(\frac{1+\left|r_{1}\right|}{2}\right)$
when $\overrightarrow{n}=(1,0,0)$. From $\rho_{B}=p_{+}\rho_{+}^{B}+p_{-}\rho_{-}^{B}=\frac{1}{2}I_2$
and $S(\rho_{B})=1$, we immediately have $C_{1}=1-h\left(\frac{1+\left|r_{1}\right|}{2}\right)$.
The basis for Alice's projection
$P_{\pm}^{A}=\frac{1}{2}(I_2 \pm\overrightarrow{n}'\cdot\overrightarrow{\sigma})$ in the definition of $Q_{2}$
must be mutually unbiased to the basis parameterized by $\overrightarrow{n}=(1,0,0)$;
therefore, the unit vector $\overrightarrow{n}'$ must be in the
form $\overrightarrow{n}'=(0,n_{2},n_{3})$. The maximum in the definition
of $Q_{2}$ is reached when $\overrightarrow{n}'=(0,1,0)$, and
thus a calculation similar to that for $C_{1}$ yields $Q_{2}=1-h\left(\frac{1+\left|r_{2}\right|}{2}\right)$.
For a qubit system, three MUBs exist. We can reveal the quantum correlation
in another (the last) MUB, which corresponds to the case $\overrightarrow{n}''=(0,0,1)$.
We easily obtain $Q_{3}=1-h\left(\frac{1+\left|r_{3}\right|}{2}\right)$.
For the general case in which the numbers $r_j$ do not follow $\left|r_{1}\right|\geq\left|r_{2}\right|\geq\left|r_{3}\right|$,
a similar argument yields $\overrightarrow{C}=(\chi_{1},\chi_{2},\chi_{3})$
with $\chi_{j}=1-h\left(\frac{1+\left| \overline{r}_{j} \right|}{2}\right)$.

%\subsection{Symmetric correlation vector}
\medskip
\noindent
{\bf Symmetric correlation vector.}
The correlation vector defined in the main text relies on a special choice
of the measure of classical correlation; it is not symmetric with
respect to exchange of A and B.
Here, we consider an alternative definition of the correlation vector, which is symmetric
with respect to the exchange of A and B.

For any bipartite quantum state
$\rho_{AB}$, Alice chooses a basis $\left\{ \left|a_{i}\right\rangle _{A}|i=1,\cdots,d_{A}\right\} $
of her system in a $d_{A}$-dimensional Hilbert space and Bob chooses
a basis $\left\{ \left|b_{i}\right\rangle _{A}|i=1,\cdots,d_{B}\right\} $
of his system in a $d_{B}$-dimensional Hilbert space, and each one
performs a measurement projecting his/her system onto the corresponding
basis states. With probability $p_{ij}=tr_{AB}((\left|a_{i}\right\rangle _{A}\left\langle a_{i}\right|\otimes\left|b_{j}\right\rangle _{A}\left\langle b_{j}\right|)\rho_{AB})$,
Alice and Bob will obtain the $i$-th and $j$-th results, respectively.
The correlation of their measurement results is well characterized
by the classical mutual information:
\begin{equation}
I\left\{ p_{ij}\right\} =H\left\{ p_{i}^{a}\right\} +H\left\{ p_{j}^{b}\right\} -H\left\{ p_{ij}\right\} ,
\end{equation}
where $H\left\{ p_{k}\right\}$ is the Shannon entropy of the probability
distribution $\left\{ p_{k}\right\} $, and $p_{i}^{a}=\sum_{j}p_{ij}$ and
$p_{j}^{b}=\sum_{i}p_{ij}$ are the marginal probability distributions.
In other words, $H\left\{ p_{i}^{a}\right\} =\sum_i - p_i^a \log _2 p_i^a$, and $H\left\{ p_{ij}\right\} = \sum_{ij} - p_{ij} \log _2 p_{ij}$.

The symmetric measure of classical correlation $C^s_{1}$ in the state $\rho_{AB}$ is defined
as the maximal classical mutual information of the local measurement
results, maximized over all local bases for both sides, i.e.,
\begin{equation}
C^s_{1}(\rho_{AB})=\max_{\{\left|a_{i}\right\rangle \otimes\left|b_{j}\right\rangle \}}I\left\{ p_{ij}\right\} .\label{defC1-1}
\end{equation}
The symmetric measure of classical correlation was discussed in \cite{wu2009correl} (where the notation $I_{max}$ was used).
A product basis $\left\{ \left|a_{i}^{1} \right\rangle  \otimes \left|b_{j}^{1}\right\rangle |i=1,\cdots,d_{A}, j=1,\cdots,d_{B} \right\} $ that achieves the maximum in (\ref{defC1-1}) is called a $C^s_1$-basis, and is denoted as
$\left\{ \left|\mathcal{A}_{i}^{1} \right\rangle \otimes \left|\mathcal{B}_{j}^{1}\right\rangle |i=1,\cdots,d_{A}, j=1,\cdots,d_{B} \right\} $.

The symmetric measure of maximal quantum correlation $Q^s_{2}$ is
defined as the maximal residual correlation over all local bases that
are mutually unbiased to a $C^s_1$-basis $\left\{ \left|\mathcal{A}_{i}^{1} \right\rangle \otimes \left|\mathcal{B}_{j}^{1}\right\rangle  \right\} $, further maximized over all possible $C^s_1$-bases (if not unique), i.e.,
\begin{equation}
Q^s_{2}(\rho_{AB})=\max_{\left\{ \left|\mathcal{A}_{i}^{1} \right\rangle \otimes \left|\mathcal{B}_{j}^{1}\right\rangle  \right\} } \max_{\{\left|a_{i}^{2}\right\rangle \otimes\left|b_{j}^{2}\right\rangle \}}I\left\{ p'_{ij}\right\} ,
\label{defQ2-1}
\end{equation}
where $\left\{ \left|a_{i}^{2}\right\rangle \right\} $ ($\left\{ \left|b_{j}^{2}\right\rangle \right\} $)
is mutually unbiased to $\left\{ \left|\mathcal{A}_{i}^{1}\right\rangle _{A}\right\} $
($\left\{ \left|\mathcal{B}_{j}^{1}\right\rangle _{B}\right\} $)
and $p'_{ij}=tr_{AB}((\left|a_{i}^2\right\rangle _{A}\left\langle a_{i}^2\right|\otimes\left|b_{j}^2\right\rangle _{A}\left\langle b_{j}^2\right|)\rho_{AB})$.
A basis $\{\left|a_{i}^{2}\right\rangle \otimes\left|b_{j}^{2}\right\rangle \}$ that achieves the maximum in (\ref{defQ2-1}) is called a $Q^s_{2}$-basis, and is denoted as
$\left\{ \left|\mathcal{A}_{i}^{2}\right\rangle \otimes\left|\mathcal{B}_{j}^{2}\right\rangle \right\} $.
We redefine the $C^s_1$-bases $\left\{ \left|\mathcal{A}_{i}^{1}\right\rangle \otimes\left|\mathcal{B}_{j}^{1}\right\rangle \right\} $
as the bases that achieve the maximum in (\ref{defC1-1}) as well as the maximum in (\ref{defQ2-1}).

Similarly, $Q^s_{3}$ denotes the residual quantum correlation in a
third complementary basis that is mutually unbiased to both $\left\{ \left|\mathcal{A}_{i}^{1}\right\rangle \otimes\left|\mathcal{B}_{j}^{1}\right\rangle \right\} $
and $\left\{ \left|\mathcal{A}_{i}^{2}\right\rangle \otimes\left|\mathcal{B}_{j}^{2}\right\rangle \right\} $.
In this manner, we have an alternative correlation
vector $\overrightarrow{C_{s}}=(C^s_{1},Q^s_{2},Q^s_{3},\cdots,Q^s_{M})$,
which is symmetric with respect to the change of A and B.

Because the Holevo
bound is an upper bound of accessible classical mutual information,
we immediately know from the above definitions that the asymmetric correlation
vector $\overrightarrow{C}$ is an upper bound of the symmetric correlation
vector $\overrightarrow{C_{s}}$ for each component, i.e., $C_{1}\geq C^s_{1}$,
$Q_{2}\geq Q^s_{2}$, $\cdots$, $Q_{M}\geq Q^s_{M}$.
It is not difficult to demonstrate that
the asymmetric correlation vector $\overrightarrow{C}$ actually coincides with the symmetric correlation vector $\overrightarrow{C_{s}}$ (i.e., $\overrightarrow{C}=\overrightarrow{C_{s}}$)
for the CQ states, the Werner states and the two-qubit states in Eq. (\ref{twoqubitsymmtricstates}).

\medskip
\noindent
{\large \bf Acknowledgements}

\noindent
The authors like to thank Caslav Brukner and Yuchun Wu for valuable discussions. S.W. acknowledges support from the NSFC via Grant 11275181, the CAS and the National Fundamental Research Program. Z.M. acknowledges support from the NSFC via Grant 11371247. Z.C. acknowledges support from the NSFC via Grant 11201427.
S.Y. acknowledges support from the National Research Foundation and Ministry of Education (Singapore) via Grant WBS: R-710-000-008-271.

\end{document}